\documentclass[aps, pre, twocolumn]{revtex4}

\usepackage{amsmath}
\usepackage[dvips]{graphicx}

\begin{document}

\title{Phase field model of premelting of grain boundaries}
\author{Alexander E.~Lobkovsky} 
\affiliation{Physics Department, Northeastern University, Boston, MA}
\author{James A.~Warren}
\affiliation{National Institute of Standards and Technology,
  Gaithersburg, MD}

\begin{abstract}
  We present a phase field model of solidification which includes the
  effects of the crystalline orientation in the solid phase.  This
  model describes grain boundaries as well as solid-liquid boundaries
  within a unified framework.  With an appropriate choice of coupling
  of the phase field variable to the gradient of the crystalline
  orientation variable in the free energy, we find that high angle
  boundaries undergo a premelting transition.  As the melting
  temperature is approached from below, low angle grain boundaries
  remain narrow.  The width of the liquid layer at high angle grain
  boundaries diverges logarithmically.  In addition, for some choices
  of model coupling, there may be a discontinuous jump in the width of
  the fluid layer as function of temperature.
\end{abstract}

\maketitle

\section{Introduction}
\label{sec:intro}

Understanding the structure of a boundary between two crystalline
grains is a prerequisite to building theories of microstructure
development and evolution.  Even in pure materials, there are many
possibilities for this structure.  Whereas a low angle grain boundary
may be thought of as an array of dislocations \cite{li:possibility},
this description is not always accurate or useful.  Several molecular
dynamics studies indicated that grain boundaries develop a layer of
disordered, amorphous material as the temperature is raised
\cite{broughton.gilmer:grain-boundary,
  keblinski.wolf.ea:self-diffusion, nguyen.ho.ea:thermal,
  schonfelder.wolf.ea:molecular-dynamics}.  Within a few degrees of
the melting temperature $T_\mathrm{m}$, grain boundaries in pure
aluminum liquefy \cite{hsieh.balluffi:experimental}.  The directly
measured width of the liquid layer was consistent with a divergence at
$T_\mathrm{m}$.  Careful calorimetry measurements as in
\cite{zhu88:_evolut_ar_ne} may be able to pin down the nature of the
divergence.  Measurements of the dihedral angle of the bicrystal in
contact with the melt in bismuth
\cite{masamura72:_absol,vold72:_behav,glicksman69:_actamet} uncovered
a discontinuous transition as a function of misorientation of tilt
boundaries near the melting point.

Indirect probes of the structure of grain boundaries also find
evidence of a structural transition.  Measurements of grain boundary
mobility \cite{author.demianczuk:effect}, shear resistance
\cite{watanabe.kimura.ea:effect} and diffusion coefficients
\cite{budke.surholt.ea:tracer} all found discontinuous jumps in these
quantities at some characteristic temperature below the melting point.

Melting at a grain boundary as well as surface melting
\cite{dash89:_surface_melting} is a particular case of a broad class
of wetting phenomena \cite{sullivan86:_wetting}.  Generally,
interfacial melting may be complete (the width of the liquid layer
diverges as $T \rightarrow T_\mathrm{m}$) or incomplete (it stays
finite) depending on the functional dependence of the free energy on
the thickness of the liquid layer.  There may also be a discontinuous
jump in the width of the liquid layer as a function of temperature.

An explicit calculation of the free energy of a liquid layer is
difficult and has been performed only in a few cases.  For example,
Kikuchi and Cahn calculated the free energy of a disordered grain
boundary within a lattice gas model
\cite{kikuchi.cahn:grain-boundary}.  They concluded that when the
interaction of the solid-liquid interfaces is short range, the
transition, if it exists, is continuous and the width of the liquid
film diverges as a logarithm of the undercooling.  Elbaum and Schick
calculated the energy of a water film on ice due to the van der Waals
dispersion forces \cite{elbaum.schick:application} and concluded that
premelting there is incomplete.

In this paper we study the modification of the phase field model of
grain boundaries \cite{kobayashi.warren.ea:continuum}.  Our model
treats solidification as well as grain boundaries within a unified
consistent framework.  Therefore, we ought to be able to describe
melting of grain boundaries.  The present authors elucidated the
general properties of this model in
Ref.~\cite{lobkovsky01:_sharp}.  There we were able to compute, in
the limit of an infinitely sharp interface, the energy, width and
mobility of model grain boundaries.

We hope to use the predictions of our phase field model near the
melting point to pin down the correct/realistic form of the model
coupling functions which offer the essential flexibility in our model.
We apply the results of \cite{lobkovsky01:_sharp} and indeed find
that grain boundaries undergo complete or incomplete continuous
premelting transition depending on the misorientation of the two
grains.  In addition, for a particular choice of the model coupling
functions, there is a discontinuous jump in the boundary width as a
function of temperature.

In the following Section \ref{sec:model} we briefly describe the phase
field model of grain boundaries (for a better discussion see
Ref.~\cite{lobkovsky01:_sharp}).  In Sec.~\ref{sec:flat} we apply
the results of \cite{lobkovsky01:_sharp} to the modified model.
We consider the structure composed of the grain boundary plus the
layer of liquid and calculate its energy and width.  We find that high
angle grain boundaries undergo a continuous premelting transition.  We
also find that for a particular choice of the model coupling
functions, there is a discontinuous jump in the width of the liquid
layer for some boundaries.  In the next Sec.~\ref{sec:mobility}, we
show that the existence of a liquid layer greatly affects the grain
boundary migration.  We summarize our findings in
Sec.~\ref{sec:conclusions}.

\section{Model}
\label{sec:model}

The phase field $\phi$ is introduced in standard treatments of the
kinetics of the liquid-solid transition \cite{langer:models-pattern}.
It measures the degree of structural disorder.  Conventionally $\phi =
1$ corresponds to a perfect solid while $\phi = 0$ corresponds to a
liquid.  In \cite{kobayashi.warren.ea:continuum} Kobayashi Warren and
Carter introduced $\eta$ which measures the degree of orientational
disorder.  We postulate that these two order parameters are
essentially the same.  Therefore the bulk free energy density
\cite{lobkovsky01:_sharp} $f(\phi)$ in the free energy $\mathcal{F}$
\begin{multline}
  \label{eq:free_energy}
  {\mathcal{F}}[\phi, \theta] = \frac{1}{\epsilon}\int_\Omega dA
  \left[f(\phi) + \frac{\alpha^2}{2} |\nabla \phi|^2 + g(\phi) \,
    s|\nabla \theta|\right. \\ \left. + h(\phi) \,
    \frac{\epsilon^2}{2} |\nabla \theta|^2\right],
\end{multline}
should be thought of as a double well with minima at $\phi = 0,1$.
Here $\theta$ is the local crystal orientation and $\alpha$,
$\epsilon$ and $s$ are positive model parameters.  The model coupling
functions $g(\phi)$ and $h(\phi)$ must vanish in the fluid $\phi = 0$
since the orientation $\theta$ has no meaning there and therefore its
gradients should not be penalized.

The present authors analyzed the gradient flow of this free energy
\begin{subequations}
  \label{eq:model}
  \begin{eqnarray}
    \label{eq:eta_eq}
    Q(\phi, \nabla \theta)\,\tau_\phi \, \frac{\partial \phi}
    {\partial t} &=& \alpha^2 \nabla^2 \phi - f_\phi - g_\phi
    \, s |\nabla\theta| \nonumber \\ && - h_\phi \,
    \frac{\epsilon^2}{2} |\nabla\theta|^2, \\ 
    \label{eq:theta_eq}
    P(\phi, \nabla\theta) \, \tau_\theta \, \phi^2 \, \frac{\partial
      \theta}{\partial t} &=& \nabla \cdot
    \left[
      h \, \epsilon^2 \nabla \theta + g \, s\,\frac{\nabla\theta}
      {|\nabla\theta|} 
    \right],
  \end{eqnarray}
\end{subequations}
in Ref.~\cite{lobkovsky01:_sharp}.  We used the subscripts to
denote differentiation except for the time constants $\tau_\eta$ and
$\tau_\theta$.  The mobility functions $P$ and $Q$ must be positive
definite, but are otherwise unrestricted.

\begin{figure}[htbp]
  \centering
  \includegraphics[width=3in,keepaspectratio]{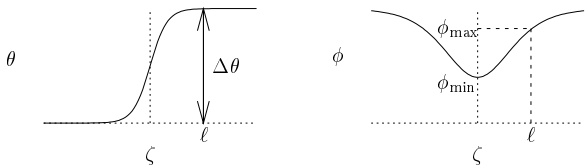}
  \caption{Order parameters in a flat stationary grain boundary as a
    function of the coordinate $\zeta$ normal to the boundary.  The
    orientation $\theta$ varies smoothly between $0$ and
    $\Delta\theta$ in the grain boundary $\zeta \in [-\ell, \ell]$ and
    is constant in the interior of the grains.  The phase field $\phi$
    approaches $1$ exponentially away from the boundary.}
  \label{fig:1d_bdry}
\end{figure}

In Ref.~\cite{lobkovsky01:_sharp} we computed the structure of the
boundary, its energy and mobility the limit of a sharp interface.  Let
us briefly summarize these results here.  The structure of the
boundary is shown in Fig.~\ref{fig:1d_bdry}.  The values of the phase
field $\phi$ in the center of the grain boundary $\phi_\mathrm{min}$
and at the edge of the grain boundary $\phi_\mathrm{max}$ are found by
solving the following two equations \cite{lobkovsky01:_sharp}
\begin{eqnarray}
  \label{eq:max_min}
  g(\phi_\mathrm{max}) & = & g(\phi_\mathrm{min}) +
  \frac{\sqrt{2f(\phi_\mathrm{min}) \,h(\phi_\mathrm{min})}}{\tilde
    s}, \\
  \label{eq:delta_theta}
  \frac{\Delta\theta}{2 \tilde \alpha \tilde s} & = & 
  \!\! \int_{\phi_\mathrm{min}}^{\phi_\mathrm{max}} \!\! 
  \frac{d\phi \,\, (g(\phi_\mathrm{max}) - g)}{h\sqrt{2f  -  \tilde s^2
  \,(g(\phi_\mathrm{max}) - g)^2/h}},
\end{eqnarray}
where $\tilde\alpha = \alpha/\epsilon$ and $\tilde s = s/\epsilon$.
Once $\phi_\mathrm{min}$ and $\phi_\mathrm{max}$ are determined, we
can compute the energy of the grain boundary $\gamma_\mathrm{gb}$ and
the amount of ``fluid'' or disordered material at the interface $W$
via
\begin{eqnarray}
  \label{eq:gamma}
  \gamma_\mathrm{gb} & = & \tilde s \Delta\theta \,
  g(\phi_\mathrm{max}) + 2\tilde\alpha \int_{\phi_\mathrm{max}}^1
  d\phi \, \sqrt{2f} \nonumber \\ && + 2 \tilde\alpha
  \int_{\phi_\mathrm{min}}^{\phi_\mathrm{max}} d\phi  \,\sqrt{2f -
    \tilde s^2(g(\phi_\mathrm{max}) - g)^2/h}, \\[2mm]
  \label{eq:W}
  W & \equiv & \int_0^\infty d\zeta \, (1 - \phi) = \tilde\alpha
  \int_{\phi_\mathrm{max}}^1 d\phi \, \frac{1 - \phi}{\sqrt{2f}}
  \nonumber \\ && + \tilde\alpha
  \int_{\phi_\mathrm{min}}^{\phi_\mathrm{max}} d\phi \,
  \frac{1 - \phi}{\sqrt{2f - \tilde s^2(g(\phi_\mathrm{max}) -
      g)^2/h}}.
\end{eqnarray}

\begin{figure}
  \centering
  \includegraphics[width=3in,keepaspectratio]{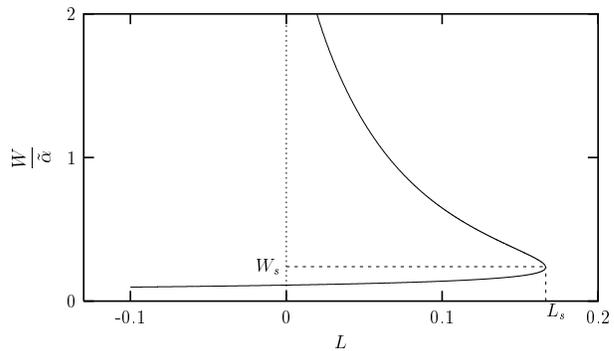}
  \caption{Amount of liquid $W$ in units of $\tilde\alpha$ a function
    of temperature $L$ for a low-angle grain boundary.  $h = \phi^2$,
    $\tilde s = 1$ and $\Delta\theta/\tilde\alpha = 0.1$.}
  \label{fig:W_low}
\end{figure}

\section{Flat stationary boundary}
\label{sec:flat}

Equipped with the way of calculating measurable properties of the
grain boundary in our model we make choices for the coupling functions
and explore the results of the calculation.  As we alluded to before
the bulk free energy density $f(\phi)$ is a double well which we
choose to be \cite{langer:models-pattern}
\begin{equation}
  \label{eq:f}
  f(\phi) = \frac{\phi^{2}(1 - \phi)^2}{2} - L \, (1 - \phi)^3(1 +
  3\phi + 6\phi^2),
\end{equation}
where $L = -f(0)$ is linear in the temperature deviation from the
melting point $T - T_\mathrm{m}$.  The choices of $g(\phi)$ and
$h(\phi)$ are more flexible.  These functions provide the essential
freedom in our model.  The only constraint on them is that they must
vanish sufficiently quickly at $\phi = 0$.  For example, they can be
chosen \cite{kobayashi.warren.ea:continuum} to match the Read-Shockley
\cite{read.shockley:dislocation} misorientation dependence of the
grain boundary energy.  In this paper we will set
\begin{equation}
  \label{eq:g}
  g(\phi) = \phi^2,
\end{equation}
and explore two choices for $h$
\begin{equation}
  \label{eq:h}
  h_1(\phi) = \phi^2 \quad \mathrm{and} \quad h_2(\phi) = 1.
\end{equation}
The second choice is problematic since it ascribes an energy cost to
changes of orientation in the liquid where it has no meaning.  We
nevertheless explore its implications to gain insight and intuition
into the role of the model function $h$.

\begin{figure}[htbp]
  \centering
  \includegraphics[width=3in,keepaspectratio]{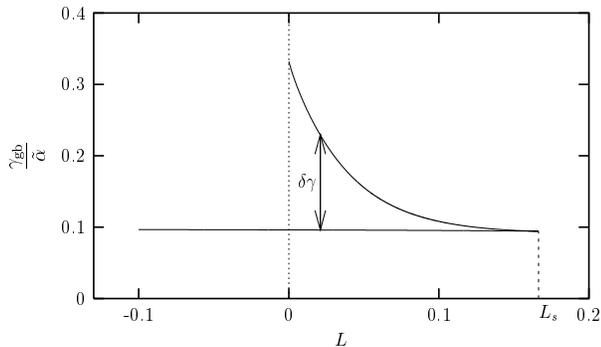}
  \caption{Energy $\gamma_\mathrm{gb}$ in units of $\tilde\alpha$ of
    the low-angle grain boundary (same values of the parameters as in
    Fig.~\ref{fig:W_low}) and the saddle point solution (upper
    branch).}
  \label{fig:gamma_low}
\end{figure}

\subsection{$h = \phi^2$}
\label{sec:h=phisq}

A typical dependence of the amount of liquid $W$ on temperature $L$
for a low-angle boundary is shown in Fig.~\ref{fig:W_low}.  The
meaning of the ``low-angle'' will become apparent in a moment.  Below
the melting temperature $L < 0$, the narrow, dry boundary is the only
solution.  Above the melting temperature, the narrow, dry boundary is
metastable with respect to the uniform fluid.  The second solution
(upper branch) is the saddle point on the path from the narrow
boundary to the globally stable uniform fluid.  It gives the size of
the critical nucleus.  The energy barrier for nucleation
$\delta\gamma$ shown in Fig.~\ref{fig:gamma_low} vanishes at the
spinodal temperature $L_s$ above which the grain boundary cannot
exist.  Note that since the width of the saddle point solution
(critical nucleus size) diverges (logarithmically
\cite{lobkovsky00:_premel_grain}) at the melting temperature, the
energy of the saddle point solution must approach that of the two
liquid solid interfaces
\begin{equation}
  \label{eq:liquid-solid}
  2\gamma_\mathrm{ls} = 2\tilde\alpha \int_0^1 d\phi \sqrt{2f} =
  \frac{\tilde\alpha}{3},
\end{equation}
which it indeed does.

\begin{figure}[htbp]
  \centering
  \includegraphics[width=3in,keepaspectratio]{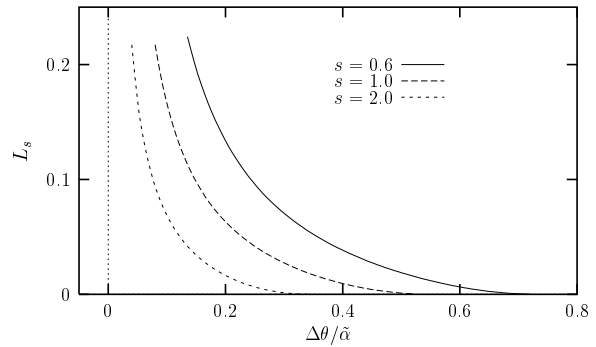}
  \caption{The spinodal temperature $L_s$, above which the grain
    boundary cannot exist, as a function of the misorientation.  Note
    that the spinodal temperature vanishes at a critical
    misorientation.}
  \label{fig:L_s}
\end{figure}

The spinodal temperature's dependence on $\tilde s$ and the
misorientation $\Delta\theta$ is shown in Fig.~\ref{fig:L_s}.  As
expected the spinodal temperature vanishes at an $\tilde s$-dependent
critical misorientation $\Delta\theta_c$.  Therefore, high-angle grain
boundaries ($\Delta\theta > \Delta\theta_c$) cannot exist above the
melting temperature.  For these boundaries, the amount of fluid $W$
diverges as the melting point is approached from below.  Another way
to define this critical misorientation is to notice that the energy
$\gamma_\mathrm{gb}$ of all high-angle grain boundaries approach that
of two liquid-solid interfaces (\ref{eq:liquid-solid}) at the melting
point, whereas low-angle grain boundaries are less energetically
costly.

\begin{figure}[htbp]
  \centering
  \includegraphics[width=3in,keepaspectratio]{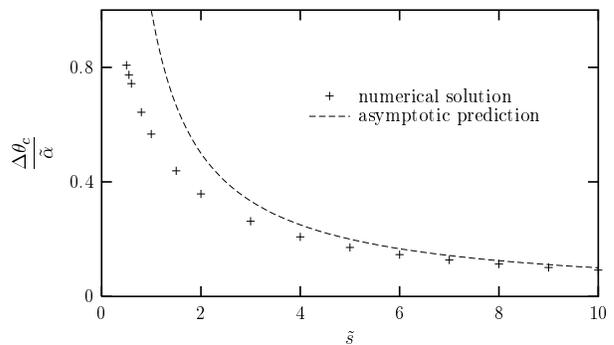}
  \caption{Critical misorientation in units of $\alpha$ as function of
    $\tilde s$.  $h = \phi^2$.  Solid line is the analytic result in
    the limit of large $\tilde s$.}
  \label{fig:delta_theta_c}
\end{figure}

The dependence of the critical misorientation on $\tilde s$, shown in
Fig.~\ref{fig:delta_theta_c}, can be calculated approximately in the
limit of large $\tilde s$.  We need to find the largest misorientation
for which a solution to (\ref{eq:delta_theta}) exists at the melting
point $L = 0$.  Using the approximate expression for the right hand
side of Eq.~(\ref{eq:delta_theta}), calculated in
\cite{lobkovsky01:_sharp}, we obtain ($\phi$ subscript denotes
differentiation)
\begin{equation}
  \label{eq:dtheta_large_s}
  \frac{\Delta\theta}{\alpha} \approx \frac{2}{\tilde
    s}\frac{\sqrt{f(\phi_\mathrm{min})}}{g_\phi(\phi_\mathrm{min})} =
  \frac{1 - \phi_\mathrm{min}}{\tilde s}.
\end{equation}
And since $\phi$ is restricted to lie within the $[0,1]$ interval, a
solution only exists for
\begin{equation}
  \label{eq:delta_theta_c}
  \Delta\theta \leq \Delta\theta_c = \frac{\tilde\alpha}{\tilde s}.
\end{equation}
Indeed, for large $\tilde s$ the numerically found critical
misorientation shown in Fig.~\ref{fig:delta_theta_c} approaches this
prediction.  It is worth noting that no solutions to our model exist
for $\tilde s < 0.5$.  In this case, some values of $\phi_\mathrm{min}
< 1$ result in unphysical $\phi_\mathrm{max}$ according to
Eq.~(\ref{eq:max_min}).

\begin{figure}[htbp]
  \centering
  \includegraphics[width=3in,keepaspectratio]{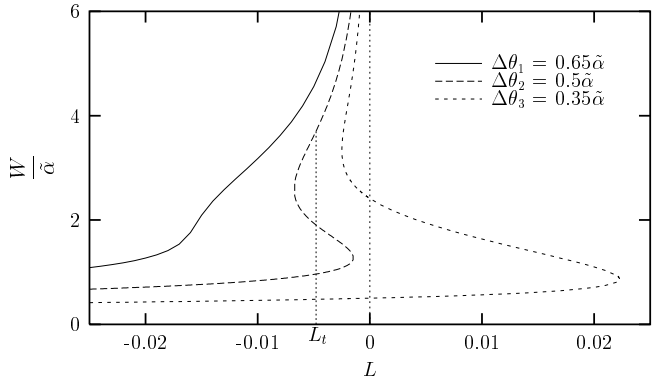}
  \caption{Amount of liquid $W$ for $h = 1$ in units of $\tilde\alpha$
    as function of temperature $L$ for $\tilde s = 1$ and three
    different misorientations $\Delta\theta_u < \Delta\theta_1$,
    $\Delta\theta_l < \Delta\theta_2 < \Delta\theta_u$ and
    $\Delta\theta_3 < \Delta\theta_l$.}
  \label{fig:W_h=1}
\end{figure}

\subsection{$h = 1$}
\label{sec:h=1}

The behavior of $W$ is more complex in this case.  As shown in
Fig.~\ref{fig:W_h=1} there are three distinct cases.  The amount of
liquid at a high-angle boundary diverges smoothly at the melting
point.  When $\Delta\theta$ is below an upper critical misorientation
$\Delta\theta_u$, there is a range of temperatures below melting in
which there are two solutions, one stable and one metastable.  For
example, by looking at the energies of these solutions, we discover,
that the narrow, dry solution is stable up to the transition
temperature $L_t < 0$ (see Fig.~\ref{fig:W_h=1}).  Above this
transition temperature, the wide, wet solution is stable, while the
narrow, dry solution is metastable.  The transition from narrow, dry
boundary to wide, wet boundary is discontinuous.

Below a lower critical misorientation $\Delta\theta_l$, there is still
a region of temperatures in which there exists a wide, wet solution,
but the narrow, dry solution is stable all the way up to the melting
temperature.

\begin{figure}[htbp]
  \centering
  \includegraphics[width=3in,keepaspectratio]{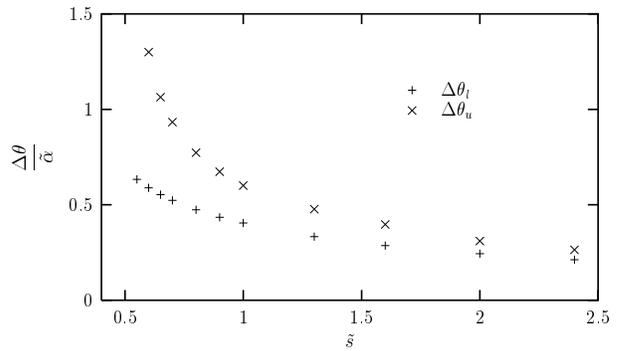}
  \caption{Upper and lower critical misorientations as a function of
    $\tilde s$ for the $h = 1$ case.}
  \label{fig:dtheta_u_and_l}
\end{figure}

This definition of the upper and lower critical misorientations lends
itself well to predicting the result of raising the temperature
slowly.  According to our definition, a low-angle grain boundary
($\Delta\theta < \Delta\theta_l$) will remain narrow up the melting
point where it will melt via a first order activated transition.
Boundaries with intermediate misorientations ($\Delta\theta_l <
\Delta\theta < \Delta\theta_u$) will experience a discontinuous jump
in width at some characteristic temperature $L_t < 0$.  When the
temperature is increased further, the width of the liquid layer will
diverge logarithmically at the melting point.  High-angle boundaries
($\Delta\theta > \Delta\theta_u$) experience a smooth divergence of
the width of the liquid layer as the melting point is approached.  The
dependence of the lower and upper critical misorientations on $\tilde
s$ is shown in Fig.~\ref{fig:dtheta_u_and_l}.

\section{Grain boundary mobility}
\label{sec:mobility}

In this paper we would like to illustrate the flexibility of our model
by demonstrating that the choice of model functions is sufficient to
reproduce a wide range of behaviors observed experimentally and
numerically.  For example, molecular dynamics simulations
\cite{upmanyu99:_misor} as well as experiments
\cite{gottstein98:_migration} reveal that the mobility of tilt grain
boundaries vanishes quickly at the zero misorientation.  The same
studies find (in two dimensions) a sharp peak in the mobility at
special (high coincident site density) misorientations at which the
energy of the boundary has a cusp.

As we showed in Ref.~\cite{lobkovsky01:_sharp}, within our model,
curved boundaries move with a normal velocity proportional to their
curvature, energy $\gamma_\mathrm{gb}$ and mobility $\mathcal{M}$
which we computed in the sharp interface limit
\begin{eqnarray}
  \label{eq:M}
  \frac{1}{\mathcal{M}} &=& \tilde\alpha \tilde s^2 \tilde\tau_\theta
  \int_{\phi_\mathrm{min}}^{\phi_\mathrm{max}} d\phi \,\, P \,
  \frac{\phi^2 (g(\phi_\mathrm{max}) - g)^2}{h^2 \sqrt{2f - \tilde
      s^2(g(\phi_\mathrm{max}) - g)^2/h}} \nonumber \\
  && + \, \frac{\tilde\tau_\phi}{\tilde\alpha}
  \int_{\phi_\mathrm{min}}^{\phi_\mathrm{max}} d\phi \,\, Q \, \sqrt{2f
    - \tilde s^2(g(\phi_\mathrm{max}) - g)^2/h} \nonumber \\
  && + \, \frac{\tilde\tau_\phi}{\tilde\alpha} \int_{\phi_\mathrm{max}}^1  
  d\phi \,\, Q \, \sqrt{2f},
\end{eqnarray}
where $\tilde \tau_\phi = \tau_\phi/\epsilon^2$ and $\tilde
\tau_\theta = \tau_\theta/\epsilon^2$.

We studied the scaling of the mobility in the limit of the vanishing
misorientation $\Delta\theta$ and found that this limit is governed by
the behavior of the model functions near $\phi = 1$.  In this paper
$g$ and $h$ are regular at $\phi = 1$.  Therefore, the behavior of $P$
and $Q$ near $\phi = 1$ determines the scaling of the grain boundary
mobility $\mathcal{M}$ for small misorientations.  If the most
strongly divergent one of the two functions $P$ and $Q$ behaves like
$(1 - \phi)^{-\mu}$, then the mobility scales like $(\Delta\theta)^{2
  - \mu}$.  As we remarked in \cite{lobkovsky01:_sharp}, the
divergence of $P$ and $Q$ at $\phi = 1$ is physically plausible since
we expect the dynamics of $\phi$ and $\theta$ to slow down in the
grain interior.

\begin{figure}[htbp]
  \centering
  \includegraphics[width=3in,keepaspectratio]{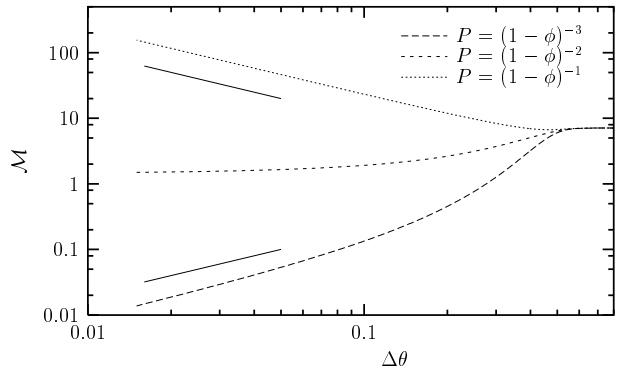}
  \caption{Grain boundary mobility $\mathcal{M}$ as a function of
    misorientation both in units of $\tilde \alpha$ for $h = \phi^2$
    and three choices of $P$.  The slopes of the two straight line
    segments are $+1$ and $-1$.  Temperature is $L = -0.001$, $\tilde
    s = \tilde\alpha = 1$.  Note that above the critical angle,
    different $P$ makes does not influence the grain boundary
    mobility.}
  \label{fig:M_theta}
\end{figure}

In Fig.~\ref{fig:M_theta} we show the grain boundary mobility as a
function of misorientation for $h = \phi^2$, $Q = -\log(1 - \phi)$ and
three different choices of $P$.  Indeed, the behavior of $P$ near
$\phi = 1$ dictates whether the mobility $\mathcal{M}$ vanishes or
diverges at zero misorientation.  The undercooling $L = -0.001$ is
small in this figure.  Larger undercoolings do not alter the
qualitative features of the situation.  Our model can therefore be
adjusted to describe systems in which the mobility dips at small
misorientations as well as systems which exhibit peaks in the mobility
at special misorientations.

Since the grain boundary mobility is strongly dependent on the amount
of disorder $W$, its behavior as a function of temperature can be
inferred from its behavior as a function of misorientation.
Specifically, when $\mu > 2$, the mobility is an increasing function
of the amount of disorder.  And vice versa $\mathcal{M}$ is a
decreasing function of $W$ when $\mu < 2$.  The amount of disordered
material at the grain boundary grows with increasing temperature as
well as with increasing misorientation.  We therefore expect the
mobility to be an increasing function of temperature when $\mu > 2$.

\begin{figure}[htbp]
  \centering
  \includegraphics[width=3in,keepaspectratio]{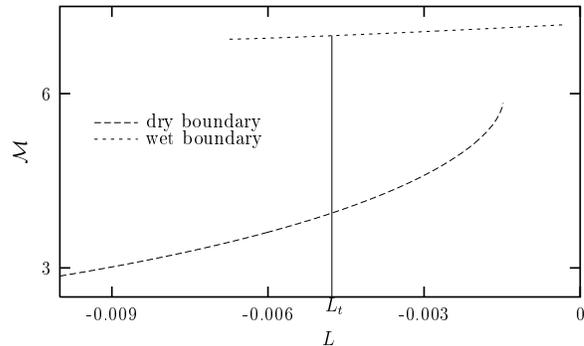}
  \caption{Mobilities of the wet and dry boundary solutions.  Here
  $\tilde s = \tilde\alpha = 1$, $\Delta\theta = 0.5\tilde\alpha$ and
  $h = \phi^2$.}
  \label{fig:M_h=1}
\end{figure}

To illustrate this point we set $h = 1$ and plot in
Fig.~\ref{fig:M_h=1} the mobility of a grain boundary which undergoes
a discontinuous premelting transition, corresponding to
$\Delta\theta_2$ in Fig.~\ref{fig:W_h=1} ($\Delta\theta_l <
\Delta\theta_2 < \Delta\theta_u$).  We also choose $\mu = 3$ so that
narrow boundaries move slower.  The mobility of the narrow, dry
boundary, which is stable below $L_t$, is significantly smaller than
that of the wide, wet boundary.  Thus when the boundary undergoes the
transition it will suddenly accelerate to almost twice the speed.

\section{Conclusions}
\label{sec:conclusions}

In this article we explored the properties of a phase field model of
crystal grains.  The model is constructed by introducing an order
parameter $\phi$ which measures the degree of disorder (or
``fluidity'') and the average local orientation $\theta$.  When the
phenomenological free energy contains a term which is linear in
$|\nabla\theta|$, its relaxational kinetics is singular and has to be
interpreted with the help of the extended gradient theory.  The
relaxing solutions thus obtained are interpreted as a collection of
grains in which $\theta$ is spatially uniform, separated by narrow
grain boundaries.

Our main goal was to demonstrate the ability of this phase field model
to reproduce experimentally observed behaviors of grain boundaries.
Since our model treats solid grain boundaries and liquid-solid
interfaces within a unified framework, it ought to correctly describe
the premelting transition of grain boundaries.  We indeed find that
depending on the choice of the model functions the grain boundary
undergoes either a continuous or a discontinuous premelting
transition.

If we choose the coefficient of $(\nabla\theta)^2$ in the free energy
to be $h(\phi) = \phi^2$, the transition is continuous.  This means
that there exists a critical misorientation $\Delta\theta_c$ such that
grain boundaries with lower misorientations remain narrow and dry all
the way up to the melting temperature.  When the misorientation is
larger than critical, grain boundaries develop a layer of fluid whose
width diverges at the melting temperature as the logarithm of the
undercooling.  The logarithmic divergence is a consequence of the
short range interaction of liquid-solid interfaces within this theory.

When $h = 1$, there is a range of misorientations for which the width
of the liquid film undergoes a jump at some characteristic temperature
below the melting temperature.  If the temperature is raised further,
the width of the liquid film diverges logarithmically as before.  Low
angle grain boundaries remain dry and narrow as before.  The jump in
the width of the fluid layer is due to the details of the short-range
interaction of the two liquid solid interfaces and should be
observable in some real systems.

Finally, we have shown that by adjusting the nature of the divergence
of the model function $P$ we are able to obtain a wide range of
behaviors of the grain boundary mobility.  In particular, depending on
the exponent of the divergence of $P$ near $\phi = 1$, the mobility of
the grain boundary can either vanish at the zero misorientation or it
can diverge.  Thus the model can emulate the behavior of the mobility
observed experimentally near the special misorientations as well as
near the zero misorientation.

In principle, by introducing appropriate additional terms in the free
energy, one can construct a more realistic model in which the grain
boundary energy has many minima at special misorientations in addition
to the cusp at the zero misorientation.  It would then be possible to
choose $P$ and $Q$ to ensure appropriate peaks or zeroes of the grain
boundary mobility at the special misorientations as well as the zero
misorientation.

\bibliography{/u/home2/leapfrog/texmf/bibtex/bib/all}

\end{document}